\documentclass[final,5p,times,twocolumn]{elsarticle}

\usepackage[T1]{fontenc}
\usepackage[utf8]{inputenc}
\usepackage{microtype}

\usepackage{amsmath,amssymb,amsfonts,mathtools}

\usepackage{graphicx}

\usepackage{pgfplots}
\pgfplotsset{compat=1.18}

\usepackage{xcolor}
\usepackage[colorlinks=true,linkcolor=blue,citecolor=blue,urlcolor=blue]{hyperref}

\journal{Physics Letters B}

\begin{document}

\begin{frontmatter}

\title{Holographic Constraints on the String Landscape}

\author[inst1]{Alek Bedroya}
\author[inst2]{Paul J.~Steinhardt}
\affiliation[inst1]{organization={Princeton Gravity Initiative},
            city={Princeton, NJ 08544},
            country={USA}}
\affiliation[inst2]{organization={Department of Physics, Princeton University},
            city={Princeton, NJ 08544},
            country={USA}}

\begin{abstract}
We show that holography imposes strong and general constraints on scalar field potentials in the string landscape, determined by the asymptotic structure of the underlying spacetime. Applying these holographic consistency conditions, we identify broad classes of scalar potentials that are incompatible with a well-defined dual description. These include potentials with extended plateaus, excessively steep or shallow asymptotics, certain zero crossings, and specific alignments of stable AdS minima in moduli space.  In particular, making the standard assumption that the CFT dual to a stable AdS vacuum must be realized as a worldvolume theory of a brane in string theory, we show that the brane selects an infinite-distance limit in moduli space where parametric scale separation is forbidden. Furthermore, the steepness and positivity of the potential are restricted in that infinite distance direction. We also find that requiring the validity of the effective theory in the future vacuum—a natural holographic criterion—automatically enforces the Trans-Planckian Censorship Conjecture (TCC) for classical cosmological solutions with positive potentials.  Taken together, these constraints exclude the leading proposals to realize scale-separated AdS vacua and metastable de Sitter vacua in the string theory landscape such as DGKT and KKLT.
\end{abstract}

\begin{keyword}
String theory \sep Holography \sep Swampland \sep Scalar potentials
\end{keyword}

\end{frontmatter}

\section{Introduction}

In string theory, infinite-distance limits of moduli space display universal behavior.  
First, scalar potentials fall off exponentially fast in weak-coupling limits at infinite distance in field space~\cite{Dine:1985he}.  
Second, an infinite tower of light, weakly coupled states emerges, with masses decreasing exponentially as functions of the scalar moduli fields in the asymptotic limi~\cite{Ooguri:2006in}.  
This implies that the lightest tower mass must scale polynomially with the potential.  
Finally, as this tower becomes lighter, gravity grows strongly coupled at a parametrically lower scale than $M_{\rm Pl}$, causing an exponential fall-off of the quantum gravity cutoff denoted by $\Lambda_s$~\cite{Dvali:2007hz,Dvali:2007wp,Dvali:2010vm,vandeHeisteeg:2022btw}.  
These features are well established in string theory and are conjectured to have generalizations that persist throughout the interior of moduli space.  

In this work we focus on two conjectures motivated by these features. One applies to negative scalar potentials, the other to positive potentials:  

\textit{Strong AdS Distance Conjecture (SADC)~\cite{Lust:2019zwm}:}  
In any AdS vacuum described by Einstein gravity, there exists a tower of states with mass $m$ satisfying
\begin{align}
    m^2 < A\,|\Lambda_{\rm AdS}|\,,
\end{align}
where $A$ is an $\mathcal{O}(1)$ constant.%

\textit{Trans-Planckian Censorship Conjecture (TCC)~\cite{Bedroya:2019snp}:}  
Any FRW phase of accelerated expansion with scale factor $a$ and Hubble parameter $H$ obeys
\begin{align}
    \frac{a_f}{a_i}\,l_P \;\leq\; \frac{1}{H_f}\,,
\end{align}
where $i$ and $f$ denote the initial and final times.

If these conjectures are correct, it is imperative to establish them under minimal assumptions, given their profound implications for the interior of moduli space, which is likely where our universe resides. This motivates developing sharp, bottom--up principles that apply away from weak--coupling boundaries. In this Letter, we draw from a series of recent works~\cite{Bedroya:2022tbh,Bedroya:2024zta,Bedroya:2025ris,Bedroya:2025ltj} to formulate new holographic methods that yield robust constraints on scalar potentials and establish special cases of the conjectures above.

The central logic is that, in theories where the scalar potential vanishes at infinite distance in field space, one can construct solutions in which the scalar itself runs off to infinity.   
Since observables in quantum gravity are defined at the boundary, the behavior of fields at infinity encodes detailed information about the interior of spacetime, and hence of moduli space. Thus, the structure of the interior of the moduli space is not independent of but, rather, highly constrained by its asymptotic behavior. Our findings are in close spirit with, and complementary to, the results of Ref.~\cite{Lust:2024aeg}, which constructed an index that encodes information about the interior of moduli space in terms of its asymptotic behavior.

\section{Constraints on scalar field potentials}

We begin by establishing four conditions (C1--C4) that any scalar potential in a consistent quantum–gravitational theory must satisfy.

\textit{C1: TCC for classical scalar cosmologies}~\cite{Bedroya:2024zta}.  
Every expanding, spatially flat FRW cosmology driven by a scalar field that runs to infinite distance without tunneling must satisfy the TCC.

\textit{C2: Asymptotic No--Scale--Separation (ANSS)}~\cite{Bedroya:2025ltj}.  
For every stable $AdS$ critical point in the landscape, there must exist an infinite--distance limit $\varphi\to\infty$ aligned with $\nabla V$ in which $V\to 0^-$ and
\begin{equation}\label{ANSS}
    \partial_\phi\!\ln V \;\partial_\phi\!\ln\Lambda_s \le \frac{2}{d-2},
\end{equation}
in Planck units.  
Equivalently,
\begin{equation}
    \frac{\nabla V}{V}\!\cdot\!\frac{\nabla\Lambda_s}{\Lambda_s}
    \le \frac{2}{d-2},
\end{equation}
where $\nabla$ is taken with respect to the light moduli and the metric on field space is $G_{IJ}$ from the kinetic term  
$\mathcal{L}_{\rm kin} = \tfrac12\, G_{IJ}\,\partial_\mu\Phi^I\partial^\mu\Phi^J$. This condition is closely tied to the relation between $V$ and the mass scale $m$ of the lightest tower of states.  
As shown in~\cite{Etheredge:2024tok}, the Emergent String Conjecture~\cite{Lee:2019wij} implies the universal relation~\cite{Castellano:2023stg,Castellano:2023jjt} 
\begin{equation}\label{eq:universal}
    \frac{\nabla m}{m}
    \!\cdot\!
    \frac{\nabla\Lambda_s}{\Lambda_s}
    = \frac{1}{d-2}.
\end{equation}
(See~\cite{Anchordoqui:2025izb} for an interesting connection of this structure to the quantum mechanics of the one–dimensional theory obtained after compactification.)
Using~\eqref{eq:universal}, the ANSS condition can be written as
\begin{equation}
    \frac{\nabla V}{V}\!\cdot\!\frac{\nabla\Lambda_s}{\Lambda_s}
    \le
    \frac{\nabla m^2}{m^2}\!\cdot\!\frac{\nabla\Lambda_s}{\Lambda_s},
\end{equation}
implying that $m^2$ must decay at least as fast as $V$ along $\nabla\Lambda_s$, hence the connection to scale separation.

\medskip
Two corollaries of the derivation of ANSS, established in this work, are:

\textit{C3: Existence of a brane.}  
There must exist a stationary warped solution with a spatially varying scalar field, i.e. the dual brane, in which the scalar field interpolates from the AdS critical point to the infinite--distance limit satisfying ANSS.

\textit{C4: Bound on potential steepness.}  
In the infinite--distance limit where $V\to 0^-$, the exponential falloff cannot be too steep:
\begin{equation}\label{ANSSFO}
    \bigl|\nabla V / V\bigr|
    \le
    2\sqrt{\frac{d-1}{d-2}}.
\end{equation}
This constraint excludes steep negative potentials from admitting a consistent holographic dual.

A partial argument for {\it C1} was given in Ref.~\cite{Bedroya:2022tbh}, showing that the TCC is automatically satisfied in string theory at the future infinity of power--law FRW backgrounds $a(t)\!\sim\!t^p$ where the scalar field $\phi$ rolls down an exponential potential.  
If TCC is violated at future infinity (i.e. $p>1$) the boundary two--point functions of the tower of light states vanish at separated points. This implies that all boundary correlators, including higher--spin modes, are trivial.  
Since no boundary observables such as scattering amplitudes or frozen correlators exist in this case, the configuration violates the holographic principle.  
The argument relies on $\lim_{t\rightarrow\infty}m/H=\infty$ and therefore does not exclude metastable de Sitter vacua where $m/H$ is finite and boundary data can be defined through dS/CFT~\cite{Strominger:2001pn}.  
Rather, it shows that after tunneling from a metastable de Sitter vacuum, the late--time expansion must satisfy $a(t)\!\sim\!t^p$ with $p\!\le\!1$, which is precisely the TCC bound~\cite{Bedroya:2019snp} for power--law FRW backgrounds.

To complete the argument for {\it C1}, Ref.~\cite{Bedroya:2024zta} proved by contradiction that the TCC holds in the interior of moduli space.  
Consider a background that transitions from an accelerating phase $a(t)\!\sim\!t^q$ with $q>1$ to a decelerating phase $a(t)\!\sim\!t^p$ with $\frac{1}{2}<p<1$.  
Using the WKB approximation to the Mukhanov--Sasaki equation, one finds unavoidable particle production in the accelerating phase,
\begin{equation}
n_k \!\sim\!
\Big(\tfrac{k}{\mathcal{V}_{\max}}\Big)^{-2\sqrt{\frac{q(2q-1)}{(1-q)^2}}
-\sqrt{\max\!\big(0,\frac{p(2p-1)}{(1-p)^2}\big)}}.
\end{equation}
At small $k$, if the accelerating phase lasts long enough to violate TCC, this production leads to trans--Planckian excitations when the vacuum is evolved backward in time, contradicting the validity of the effective theory in the future vacuum.  
Hence, any scalar cosmology that rolls to infinite distance without tunneling must satisfy the TCC.

As for {\it C2}, ANSS: Holographic duals of AdS vacua arise when low–energy degrees of freedom localized on branes (or in near–horizon regions of black branes) \emph{decouple} from bulk gravity. 
Upon dimensional reduction on the transverse directions, extremal black branes take the form
\begin{equation}\label{eq:DC-metric}
  ds^2=a(x)^2\!\left(-dt^2+\sum_{i=1}^{d-2} d\sigma_i^2\right)+dx^2,
\end{equation}
where $x\!\to\!-\infty$ corresponds to the AdS throat and $x\!\to\!+\infty$ to the asymptotic region. 
The scalars freeze at their AdS values in the throat but run to infinite distance as $x\!\to\!\infty$—for instance, the $S^5$ volume modulus in $AdS_5\times S^5$ diverges away from the brane.

In this background, local energies redshift as $E(x)\!\propto\!a(x)^{-1}$.
Decoupling requires the redshifted energies to remain below the local quantum–gravity cutoff $\Lambda_s(x)$ (the species scale),
\begin{equation}\label{eq:ggrav}
  g_{\rm grav}=\lim_{x\to\infty}\frac{M_{\rm pl}}{a(x)\Lambda_s(x)}<\infty.
\end{equation}
If $g_{\rm grav}\!\to\!\infty$, gravity becomes infinitely strongly coupled at the boundary and EFT breaks down.

If, say,  the potential away from the brane has exponential fall–off 
$V(\phi)=V_0e^{-\lambda\phi}$, then the solutions are 
\begin{align}
\lambda \ge \lambda_c:&\quad 
  a(x)\!\sim\!x^{1/(d-1)},\quad 
  \phi(x)\!\sim\!\sqrt{\tfrac{d-2}{d-1}}\log x,\\
V_0<0~\&~\lambda < \lambda_c:&\quad 
  a(x)\!\sim\!x^{4/[(d-2)\lambda^2]},\quad 
  \phi(x)\!\sim\!\tfrac{2}{\lambda}\log x,\\
V_0>0~\&~\lambda < \lambda_c:&\quad 
  \rm{singular},
\end{align}
with $\lambda_c=2\sqrt{(d-1)/(d-2)}$. Let $c\!\equiv\!|\partial_\phi\Lambda_s|/\Lambda_s$ be the logarithmic derivative of the species scale along $\nabla V$.  Suppose ANSS~\eqref{ANSS} is violated and 
\begin{equation}\label{eq:c-bound}
  \lambda\cdot c \;=\; \frac{2+\epsilon}{(d-2)}.
\end{equation}
for some positive $\epsilon$.

\paragraph*{Case I: ${\lambda>\lambda_c}$.}
From the asymptotics,
\begin{equation}
  \frac{M_{\rm pl}}{a(x)}\!\sim\!x^{-1/(d-1)}\!\sim\!
  \exp\!\!\Big[-\frac{\phi}{\sqrt{(d-1)(d-2)}}\Big],
\end{equation}
and the Emergent String Conjecture implies $\Lambda_s$ decreases at least this fast, with saturation only for one–dimensional decompactification~\cite{vandeHeisteeg:2023uxj}. 
Thus $g_{\rm grav}\!\to\!\infty$ unless exactly one extra dimension decompactifies, corresponding to codimension–2 branes where a large–$N$ limit is obstructed by a conical deficit.

\paragraph*{Case II: $V_0<0~\&~{\lambda\le\lambda_c}$.}
Here
\begin{equation}
  \frac{M_{\rm pl}}{a(x)}\!\sim\!
  x^{-4/[(d-2)\lambda^2]}\!\sim\!
  \exp\!\!\Big[-\frac{2\phi}{(d-2)\lambda}\Big],
\end{equation}
and using \eqref{eq:c-bound},
\begin{equation}
  g_{\rm grav}\!\sim\!
  \exp\!\!\Big[\Big(c-\frac{2}{(d-2)\lambda}\Big)\phi\Big]\!\to\!\infty,
\end{equation}
and therefor the modes on the brane cannot decouple again.

\paragraph*{Case III: ${V_0>0~\&~\lambda\le\lambda_c}$.}
If $V_0>0$ and $\lambda < \lambda_c$, the scalar field will be drawn to the local maximum of the scalar potential and the spacetime will become singular. So this possibility is rulled out as well. 

Therefore, in all three cases, the decoupling test fails. Hence, we must have $\lambda\cdot c\leq 2/(d-2)$ which implies ANSS is satisfied. Otherwise, perturbative modes exceed the species scale asymptotically and gravity cannot decouple from any mode in the theory.
This condition is the spacelike analog of the time--dependent argument in~\cite{Bedroya:2025ris}.

Furthermore, the above reasoning automatically implies {\it C3} and {\it C4}. The very existence of a dual brane which has a decoupling limit is equivalent to {\it C3}. Similarly, there follows {\it C4},
a bound on the steepness of potentials: 
Cases~I and III above, corresponding to 
$|\nabla V/V|>2\sqrt{(d-1)/(d-2)}$ or $V_0>0$, 
never yield a decoupled brane with a valid large--$N$ limit admitting an Einstein--gravity dual, regardless of how $\Lambda_s$ behaves. Thus, the scalar potential in the direction satisfying the ANSS must be negative and cannot be arbitrarily steep so that it belongs to Case II. This also ensures that, sufficiently far from the brane, the scalar field in the warped solution continues to follow the gradient direction, since the potential remains non-negligible in that region.

\section{Implications for Scalar Field Potentials}

We now apply the constraints derived above to identify classes of scalar potentials that are inconsistent with holographic constraints reviewed in the previous section. 
We begin with single–field potentials in four spacetime dimensions and comment later on how the inclusion of multiple light scalars modifies these results.

\subsection*{Asymptotic behavior for positive potentials}

We first consider positive scalar potentials relevant for cosmology. 
As shown in the previous section, in the infinite–distance limit of field space such potentials must exhibit a sufficiently steep exponential fall–off such that the future infinity is null and that the TCC is satisfied. 
\begin{equation}
  V(\phi) \sim e^{-\lambda\phi}, \qquad 
  \lambda \ge \sqrt{2},
\end{equation}
so that no power–law acceleration persists at late times. This condition is required by the existence of non-trivial holographic observables. 
Therefore, single–field landscapes containing potentials of the form 
$V(\phi)=V_0 e^{-\lambda\phi}$ with $\lambda<\sqrt{2}$ are ruled out.

\subsection*{Plateaus}

In addition to the constraint on the asymptotic behavior, we showed that any classical FRW solution in which the scalar field trajectory connects the interior of moduli space to its asymptotic region must also obey the TCC. 
This excludes long plateau potentials that extend directly to infinity. 
As an illustrative example, consider
\begin{equation}
  V(\phi)
  =V_0\!\left[1-\!\left(1+\tfrac{1}{2}e^{-2\sqrt{2/3}\phi}\right)^{-1}\right]\,.
\end{equation}

A trajectory starting from $\phi_i\!\ll\!-1$ undergoes a slow-roll quasi-de~Sitter phase but inevitably violates the TCC before rolling to infinity.

The number of $e$–folds is approximately
\begin{equation}
  N \simeq \tfrac{3}{4}e^{\sqrt{2/3}|\phi_i|}.
\end{equation}
TCC then imposes
\begin{equation}
  N < \ln\!\Big(\frac{M_{\rm pl}^2}{\sqrt{V_0}}\Big),
  \qquad\Rightarrow\qquad
  |\phi_i| < \ln\!\Big[\ln\!\Big(\frac{M_{\rm pl}^2}{\sqrt{V_0}}\Big)\Big],
\end{equation}
up to additive numerical factors. 
Hence, extended plateaus that connect smoothly to the asymptotic region of field space are ruled out. 
Any TCC–violating plateau must instead terminate in a metastable de Sitter critical point of the potential.

In the presence of multiple light scalars, the same conclusion holds: 
if there exists any trajectory that bypasses a metastable point and runs off to infinity, the potential is again ruled out.

\subsection*{Asymptotic behavior for negative potentials}

We can also obtain constraints for negative potentials. 
Specifically, if the landscape admits a stable AdS solution, then we must have $V\!\to\!0^-$ and $\partial_\phi\!\ln V \;\partial_\phi\!\ln\Lambda_s \le 2/(d-2)$ in an asymptotic direction which is along $\nabla V$. 
Potentials that violate this condition are inconsistent with holographic decoupling.
For instance, consider
\begin{equation}
 V_{\rm DGKT}(\phi)
 =10^4 e^{-\sqrt{\frac{{26}}{{3}}}\phi}
  - e^{-3\sqrt{\frac{{6}}{{13}}}\phi},
\end{equation}
a simplified single–field potential that captures the asymptotic behavior of the
DGKT construction (named after its authors~\cite{DeWolfe:2005uu}) for flux quanta $N\!\sim\!100$ along the
gradient–flow direction of the scalar potential (see Ref.~\cite{Bedroya:2025ltj}
for the reduction from the multi–field potential to a single–field model).

In the large–$\phi$ limit, the species scale given by the string mass and behaves as 
$\Lambda_s\!\sim\! e^{-\sqrt{\frac{{6}}{13}}\phi}$, implying 
$\partial_\phi\!\ln V \;\partial_\phi\!\ln\Lambda_s >1$ and thus violating the \emph{Asymptotic No–Scale–Separation} condition. 
Consequently, this scalar potential is ruled out by holography.

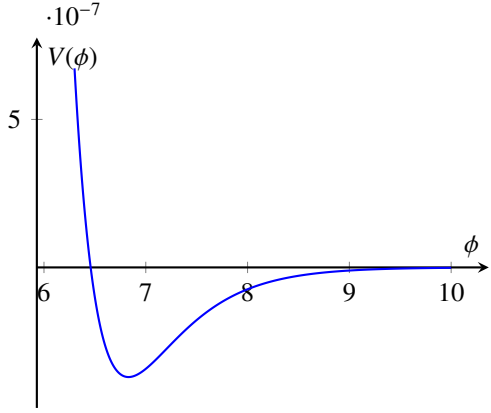
\begin{figure}[h]
\centering
\begin{tikzpicture}
\begin{axis}[
    domain=6.3:10,
    samples=400,
    xlabel={$\phi$},
    ylabel={$V(\phi)$},
    axis lines=middle,
    thick,
    no markers,
    grid=none,
    enlargelimits=true,
    width=0.85\columnwidth,
]
\addplot[blue, thick]
{1e4*exp(-(sqrt(36)/sqrt(3))*x) - exp(-(3*sqrt(6)/sqrt(13))*x)};
\end{axis}
\end{tikzpicture}
\caption{
$V_{\rm DGKT}(\phi)
 =10^4 e^{-\sqrt{\frac{{26}}{{3}}}\phi}
  - e^{-3\sqrt{\frac{{6}}{{13}}}\phi}$.
At large $\phi$, the potential violates the
Asymptotic No–Scale–Separation condition, and is therefore excluded by holography.}
\end{figure}

In the previous section, we also derived an additional constraint {\it{C4}} that depends solely on the fall--off rate of the scalar potential. 
Specifically, a negative scalar potential must possess at least one infinite--distance direction (corresponding to the black--brane solution) in which Eq.~\eqref{ANSSFO} is satisfied. This condition implies that if the scalar potential is effectively one--dimensional and admits a stable AdS critical point, its asymptotic decay rate must satisfy the same inequality, ensuring the existence of a consistent holographic brane dual. 

\subsection*{Zero crossing}

Another feature that is ruled out occurs when a one–dimensional landscape with a stable AdS vacuum has a scalar potential approaches zero from the positive direction. 
To admit a black–brane solution with a worldvolume dual to the AdS, the potential must instead approach zero from the negative side, as required by the Asymptotic No–Scale–Separation condition.

For example, 
\begin{equation}
  V(\phi) = V_0\!\left(e^{-10\phi} - 2e^{-4\phi} + e^{-2\phi}\right),
\end{equation}
crosses zero and then grows positive at large $\phi$. 
Such a potential violates the asymptotic condition and is therefore excluded by holography.

\begin{figure}[h]
\centering
\begin{tikzpicture}
\begin{axis}[
    domain=-0.05:3,
    samples=400,
    xlabel={$\phi$},
    ylabel={$V(\phi)/V_0$},
    axis lines=middle,
    enlargelimits=true,
    width=0.85\columnwidth,
    grid=none
]
\addplot[blue, thick]{exp(-10*x) - 2*exp(-4*x) + exp(-2*x)};
\end{axis}
\end{tikzpicture}
\caption{
Example of a ruled–out potential which crosses zero and approaches zero from the positive side, violating the Asymptotic No–Scale–Separation condition.
}
\end{figure}
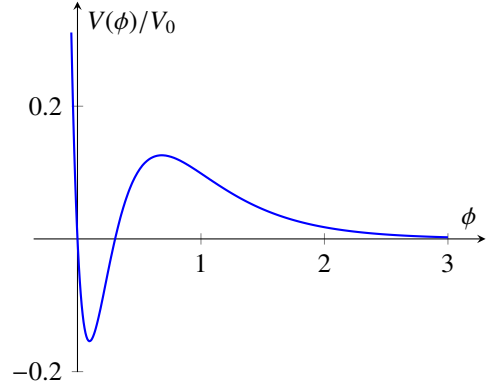

\subsection*{Multiple minima}

\begin{figure}[h]
\centering

\begin{tikzpicture}
\begin{axis}[
    domain=0.6:8,
    samples=400,
    thick,
    axis lines=middle,
    enlargelimits=true,
    grid=none,
    width=0.85\columnwidth,
    height=0.45\columnwidth,
    xtick=\empty,
    ytick=\empty,
    xlabel={$\phi$},
    ylabel={$V(\phi)$},
    ylabel style={yshift=1.2em}, 
]
\addplot[blue, thick]
{10*(exp(-4*(x-1)) - 2*exp(-2*(x-1))) - 5/cosh((x-4)*2)};
\draw[dashed, gray] (axis cs:1,0) -- (axis cs:1,{10*(exp(-4*(1-1)) - 2*exp(-2*(1-1))) - 5/cosh((1-4)*2)});
\draw[dashed, gray] (axis cs:4,0) -- (axis cs:4,{10*(exp(-4*(4-1)) - 2*exp(-2*(4-1))) - 5/cosh((4-4)*2)});
\node[above, red] at (axis cs:1,0) {$\phi_1$};
\node[above, red] at (axis cs:4,0) {$\phi_2$};
\end{axis}
\end{tikzpicture}

\vspace{1cm}

\begin{tikzpicture}
\begin{axis}[
    domain=0.6:8,
    samples=400,
    thick,
    axis lines=middle,
    enlargelimits=true,
    grid=none,
    width=0.85\columnwidth,
    height=0.45\columnwidth,
    xtick=\empty,
    ytick=\empty,
    xlabel={$\phi$},
    ylabel={$V_E(\phi)=-V(\phi)$},
    ylabel style={yshift=1.2em}, 
]
\addplot[blue, thick]
{-(10*(exp(-4*(x-1)) - 2*exp(-2*(x-1))) - 5/cosh((x-4)*2))};
\draw[dashed, gray] (axis cs:1,0) -- (axis cs:1,{-(10*(exp(-4*(1-1)) - 2*exp(-2*(1-1))) - 5/cosh((1-4)*2))});
\draw[dashed, gray] (axis cs:4,0) -- (axis cs:4,{-(10*(exp(-4*(4-1)) - 2*exp(-2*(4-1))) - 5/cosh((4-4)*2))});
\node[below, red] at (axis cs:1,0) {$\phi_1$};
\node[below, red] at (axis cs:4,0) {$\phi_2$};
\end{axis}
\end{tikzpicture}

\caption{
Illustration of a scalar potential $V(\phi)$ (top) and its Euclidean counterpart $V_E(\phi)=-V(\phi)$ (bottom). 
The two minima $\phi_1$ and $\phi_2$ correspond to AdS vacua in the Lorentzian potential, which appear as maxima in the Euclidean potential $V_E$. 
This correspondence clarifies the relation between CDL instantons and black--brane domain walls.
}
\label{fig:VandVE}
\end{figure}
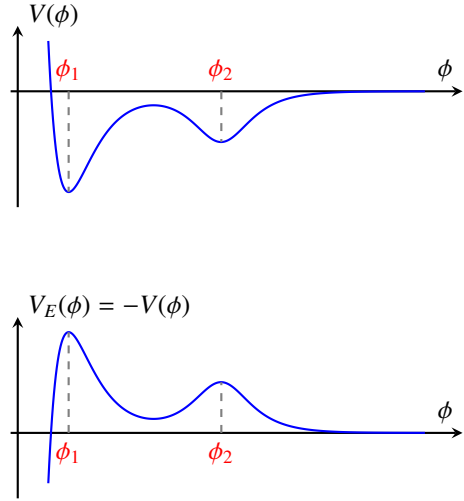

Consider a landscape with multiple AdS vacua in a single scalar potential.  
As discussed in the previous section, for each stable AdS vacuum there must exist a black--brane solution whose near--horizon geometry is that AdS, with the scalar field fixed at its critical value in the throat and flowing to infinity away from the brane.  
We show that if the landscape contains only one light scalar field, this requirement implies that two stable AdS vacua cannot coexist.  
More precisely, the stability of the AdS vacuum closer to the infinite--distance limit (labeled $\phi=\phi_2$ in Fig.~\ref{fig:VandVE}) prevents the existence of a scalar trajectory connecting $\phi=\phi_1$ to infinity.  

The condition for the stability of an AdS minimum against Coleman--De~Luccia (CDL) tunneling into a deeper vacuum is the absence of a Euclidean solution of the form
\begin{equation}
    ds^2 = d\rho^2 + a_{CDL}(\rho)^2 d\Omega_{d-1}^2,
\end{equation}
where $d\Omega_{d-1}^2$ denotes the metric on the $(d\!-\!1)$--sphere.  
The boundary conditions are $a_{CDL}(0)=0$, $a_{CDL}'(0)=1$, and $\phi_{CDL}(\rho\to\infty)=\phi_2$, while $\phi_{CDL}(0)$ lies near the first critical point $\phi_1$ with $\phi_{CDL}'(0)=0$.  
The corresponding equations of motion are
\begin{align}
    \phi_{CDL}'' + 3\frac{a_{CDL}'}{a_{CDL}}\phi_{CDL}' - V' &= 0, \\
    3\!\left(\frac{a_{CDL}'}{a_{CDL}}\right)^2 &= 3a_{CDL}^{-2} - V + \tfrac{1}{2}\phi_{CDL}'^2.
\end{align}

These Euclidean equations are equivalent to those describing a negatively curved FRW cosmology with potential $V_E=-V$.  
Similarly, the black--brane equations in the metric~\eqref{eq:DC-metric} take the form
\begin{align}
    \phi_{\mathrm{brane}}'' + 3\frac{a_{\mathrm{brane}}'}{a_{\mathrm{brane}}}\phi_{\mathrm{brane}}' + V_E' &= 0, \\
    3\!\left(\frac{a_{\mathrm{brane}}'}{a_{\mathrm{brane}}}\right)^2 &= V_E + \tfrac{1}{2}\phi_{\mathrm{brane}}'^2,
\end{align}
which are identical to those of a spatially flat FRW cosmology driven by $V_E$.  

The CDL instanton thus corresponds to a cosmological solution with negative spatial curvature, starting at $\phi>\phi_1$ at $t=0$ and evolving toward $\phi_2$ as $t\to\infty$.  
If we adjust the initial condition closer to $\phi_1$, the scalar begins with higher energy and overshoots $\phi_2$ at finite time.  
As the starting point approaches the local maximum $\phi_1$ of $V_E$, the solution has a longer period of quasi--de~Sitter expansion in the beginning. Therefore, the curvature becomes exponentially diluted by the time $\phi$ reaches $\phi_2$ and the overshoot criterion becomes curvature--independent.  
Therefore, the $\phi_2$ vacuum is unstable if and only if the spatially flat cosmological solution in the potential $V_E$ beginning at $\phi_1$ overshoots $\phi_2$---precisely the condition for the existence of a black--brane solution.  

Hence, the following statements are equivalent:
\begin{itemize}
    \item The AdS vacuum at $\phi=\phi_2$ is unstable to CDL tunneling into the vacuum at $\phi=\phi_1$.
    \item A domain--wall (black--brane) solution of the form~\eqref{eq:DC-metric} exists that asymptotes to $\phi=\phi_1$ near the horizon and flows toward $\phi\!\to\!\infty$.
\end{itemize}

If both minima were truly stable AdS vacua, the corresponding domain--wall solution would be obstructed: 
a scalar field starting from one minimum could not reach infinity because it would be trapped by the other.  
Therefore, in a one--dimensional landscape, multiple stable AdS vacua are inconsistent with holography and the existence of the required black--brane solutions.

\section{Multi--Field Generalizations}

In this section, we discuss the generalization of the results from the previous section to multi--field moduli spaces and how the corresponding constraints are strengthened or relaxed.

\subsection*{Positive potentials}

For the asymptotic behavior of positive scalar potentials, any weakly--coupled limit of the moduli space must satisfy
\begin{equation}
    \Big|\frac{\nabla V}{V}\Big| \ge \frac{2}{\sqrt{d-2}},
\end{equation}
to ensure the existence of well--defined holographic observables at future infinity.  

Regarding plateaus in the interior of moduli space that admit TCC--violating cosmological trajectories connecting to the asymptotic region, such configurations are ruled out even in the multi--field landscape.  
However, our argument based on the validity of the EFT in the future vacuum admits a loophole:  
a plateau may persist if every cosmological trajectory on it terminates in a metastable de~Sitter vacuum.  
Consequently, any TCC--violating plateau (if it exists at all) must be arranged in a highly fine--tuned manner relative to metastable de~Sitter minima, whose existence in the landscape remains itself uncertain.  
Achieving both features is already highly nontrivial in a single--field setting.  
In the generic multi--field case, where the potential depends on many moduli, these conditions are even more constraining since every trajectory in the muti-field scalar field space that goes to the asymptotics (which now can be much more complicated) must satisfy TCC. Therefore, the addition of scalar fields make the situation even more intricate and unlikely to exist.

\subsection*{Negative potentials and AdS vacua}

For every stable AdS vacuum in a negative scalar potential, holography requires the existence of an associated infinite--distance limit in field space aligned with $\nabla V$.  
This direction corresponds to the asymptotic region of the extremal black--brane solution whose near--horizon geometry is the given AdS vacuum.  
Along this direction, the following conditions must simultaneously hold:
\begin{align}
    &V \;\rightarrow\; 0^-, \quad \frac{\nabla V}{V}\cdot\frac{\nabla\Lambda_s}{\Lambda_s} \leq \frac{2}{d-2}, 
    \quad \Big|\frac{\nabla V}{V}\Big| \le 2\sqrt{\frac{d-1}{d-2}},
    \label{eq:neg-pot-conds}
\end{align}
where $\Lambda_s$ is the species scale.  
The first two conditions encode the Asymptotic No--Scale Separation (ANSS) condition, while the steepness bound reflects the requirement that the black--brane solution possesses a decoupling limit with a controlled large--$N$ dual.

These constraints are remarkably powerful when applied to multi--field potentials.  
In particular, they exclude the AdS vacua of the KKLT construction~\cite{Kachru:2003aw}, provided one assumes that the dual CFT must arise as the worldvolume theory of a brane in string theory.  
This assumption is natural for supersymmetric AdS vacua, for which all known dual SCFTs arise from branes with the corresponding near--horizon AdS geometry.

Let us briefly review some key aspects of the KKLT potential. The KKLT scenario begins with type IIB string theory compactified on a Calabi--Yau orientifold with background fluxes.  
At tree level, the four--dimensional scalar potential is 
\begin{equation}
    V_{\text{tree}}
    = e^K\!\left(K^{i\bar j} D_i W\, D_{\bar j}\overline{W} - 3|W|^2\right),
    \label{eq:Vtree}
\end{equation}
where the flux superpotential $W$ depends only on the axio--dilaton and the complex--structure moduli.  
Labeling the Kähler moduli by indices $\alpha,\beta$ and the complex--structure moduli by $a,b$, we have
\begin{equation}
    \partial_\alpha W = 0, 
    \qquad
    K^{\alpha\beta} K_\alpha K_\beta = 3
    \quad\text{(no--scale identity)}.
\end{equation}
Using this in \eqref{eq:Vtree} eliminates the $-3|W|^2$ term, leaving
\begin{equation}
    V_{\text{tree}}
    = e^K\, K^{a\bar b}\, D_a W\, D_{\bar b}\overline{W}
    \;\ge\; 0,
    \label{eq:Vtree-positive}
\end{equation}
a manifestly non--negative potential.  
Thus, at tree level:
\begin{itemize}
    \item the complex--structure moduli are stabilized at a positive semi--definite potential,
    \item the Kähler moduli remain completely flat directions.
\end{itemize}

KKLT then adds nonperturbative contributions to the superpotential,
\begin{equation}
    W_{\text{np}} \sim A\, e^{-a T},
\end{equation}
where $T$ is a Kähler modulus and itself is exponential in the canonically normalized moduli.  
These terms stabilize the Kähler sector and generate a supersymmetric AdS vacuum with
\begin{equation}
    V_{\text{KKLT}} \;=\; V_{\text{tree}} + V_{\text{np}}
    \;<\; 0.
\end{equation}

Since the nonperturbative potential behaves as a \emph{double exponential} in terms of canonically normalized moduli, in any infinite--distance limit where the non-perturbative contribution dominates we have,
\begin{equation}
    \Big|\frac{\nabla V_{\text{np}}}{V_{\text{np}}}\Big|
    \;\longrightarrow\; \infty.
    \label{eq:double-exp-steep}
\end{equation}

Consider any infinite--distance limit in the multi--field moduli space in which $V \to 0$.  
Two possibilities arise:

\paragraph*{1. The perturbative piece dominates.}
Then,
\begin{equation}
    V \;\approx\; V_{\text{tree}} \;\ge\; 0.
\end{equation}
But holography requires $V\to 0^-$ along the black--brane direction.  
Hence, perturbative domination violates ANSS.

\paragraph*{2. The nonperturbative piece dominates.}
Then the potential decays double exponentially, and
\begin{equation}
    V \;\approx\; V_{\text{np}} \;<\; 0,
\end{equation}
but the steepness becomes infinite:
\begin{equation}
    \Big|\frac{\nabla V}{V}\Big|
    \;\approx\; \Big|\frac{\nabla V_{\text{np}}}{V_{\text{np}}}\Big|
    \;\to\;\infty.
\end{equation}
This violates the steepness bound
\begin{equation}
    \Big|\frac{\nabla V}{V}\Big|
    \le 2\sqrt{\frac{d-1}{d-2}}.
\end{equation}
Thus, nonperturbative domination violates the steepness bound.

Therefore, we conclude that every infinite--distance limit of KKLT violates at least one of the holographic constraints \eqref{eq:neg-pot-conds}. Therefore, KKLT does \emph{not} admit a holographic CFT dual as the worldvolume theory of a brane in string theory.

A closely related but distinct concern was raised in Ref.~\cite{Lust:2022lfc}, where it was shown that the standard proposal for a CFT dual to KKLT would not lead to a valid low–energy gravitational description. In contrast, our argument does not rely on any particular proposal for the CFT dual. Instead, it uses only the asymptotic behavior of the scalar potential in finite–distance limits of moduli space, which is under robust control.

\subsection*{Multiple AdS minima}

Finally, let us comment on the multiplicity of AdS vacua.  
While we showed that multiple stable AdS vacua cannot coexist in a single--field potential, this restriction can be relaxed in multi--field settings.  
Indeed, an explicit example such as M--theory compactified on squashed $S^7$~\cite{Awada:1982pk} exhibit multiple stable AdS vacua.  
In such cases, each AdS minimum can independently connect to its own infinite--distance direction through a distinct black--brane solution.  
Geometrically, this means that the stable AdS vacua are not aligned along the same direction in moduli space that governs the black--brane flow.  
For instance, in M--theory on $S^7$, the different vacua correspond to distinct shape moduli of the internal manifold and are therefore not coaligned with the volume modulus of $S^7$.  
The latter provides the infinite--distance direction associated with the black--brane solution, as the internal volume diverges away from the brane.

\section{Conclusions}

By studying solutions in which the scalar field evolves from the interior toward the asymptotic regions of moduli space, we derived sharp constraints that connect the boundary behavior of the theory to the detailed shape of the potential in the bulk.  
Applying these holographic consistency conditions, we identified several conditions that must be satisfied:
\begin{itemize}
    \item For positive potentials, the potential must be steep in the asymptotics regime, i.e. $|\nabla V/V| \geq  2/\sqrt{d-2}$.
    \item For positive potentials, long plateaus are forbidden if there exists a TCC-violating classical trajectory that connects the plateau to an infinite distance region in the moduli space.
    \item For negative potentials, there must be a direction along $\nabla V$ in which three conditions are simultaneously satisfied: (i) $V$ remains negative, (ii) $\partial_\phi\ln(V)\partial_\phi\ln(\Lambda_s)\leq 2/(d-2)$, and (iii) $|\nabla V/V| \le 2\sqrt{(d-1)/(d-2)}$. For example, this condition forbids DGKT and KKLT energy landscapes.
    \item For negative potentials, having multiple stable AdS vacua along the same moduli--space direction as the black--brane flow is forbidden. 
\end{itemize}
These remarkably powerful diagnostics identify inconsistencies in low--energy effective potentials based solely on their asymptotic behavior, without requiring detailed string embeddings.  
In particular, the failure of scale separation in Ricci--flat compactifications emerges as a direct holographic consequence of the absence of internal curvature contributions to $V \sim m^2$.

Our assumptions are minimal and physically well--motivated.  
For positive potentials, we assumed the validity of the effective field theory in the future vacuum. This is a natural requirement in string theory, where the EFT is defined as an approximation to the S--matrix of a finite particle Hilbert space.  
For negative potentials, we assumed that the CFT dual to any stable AdS vacuum must arise as the worldvolume theory of a brane in string theory. This is a well–motivated assumption as string theory is widely used to classify SCFTs under the working hypothesis that all consistent SCFTs admit a realization in string theory, which then provides their ultraviolet completion (see~\cite{Heckman:2013pva,Heckman:2015bfa,Jefferson:2018irk} for examples).

A key lesson from our analysis is that these constraints apply to the \emph{entire} landscape, not to isolated regions of moduli space. Holography ties the structure at infinite distance to the physics in the interior, so consistency cannot be checked locally. For example, the existence of a single stable AdS vacuum requires a flow to an infinite-distance limit with controlled species scale and bounded potential gradient; if any asymptotic region fails to satisfy these conditions, the putative vacuum is inconsistent. The same logic excludes plateaus that would violate the TCC if there exists \emph{any} classical trajectory from the plateau to infinity. Thus, the constraints we derive are genuinely global: stability, asymptotics, and the behavior of light towers are correlated across moduli space. These results show that the interior of moduli space is tightly constrained by its asymptotic structure and provide concrete holographic diagnostics for identifying consistent scalar potentials and sharpening the boundary between the landscape and the swampland.

\section*{Acknowledgments}
We thank Fien Apers, Manki Kim, Juan Maldacena, Miguel Montero, George Tringas, Cumrun Vafa, and Timm Wrasse for valuable and helpful conversations. AB is supported in part by the Simons Foundation under grant number~654561 and by the Princeton Gravity Initiative at Princeton University. PJS is supported in part by the U.S.\ Department  of Energy under grant number~DE-FG02-91ER40671 and by the Simons  Foundation under grant number~654561.

\bibliographystyle{elsarticle-num}
\bibliography{References}

\end{document}